# Spread Spectrum based Robust Image Watermark Authentication

T. S. Das[1], V. H. Mankar[2] and S. K. Sarkar[3]

[1] Gurunanak Institute of Technology, Kolkata, [2, 3] Jadavpur University, Kolkata, [1]tirthasankardas@yahoo.com, [2]vijaymankar@yahoo.com, [3]su_sircir@yahoo.co.in

**Abstract**-In this paper, a new approach to Spread Spectrum (SS) watermarking technique is introduced. This problem is particularly interesting in the field of modern multimedia applications like internet when copyright protection of digital image is required. The approach exploits two-predecessor single attractor (TPSA) cellular automata (CA) suitability to work as efficient authentication function in wavelet based SS watermarking domain. The scheme is designed from the analytical study of state transition behaviour of non-group CA and the basic cryptography/encryption scheme is significantly different from the conventional SS data hiding approaches. Experimental studies confirm that the scheme is robust in terms of confidentiality, authentication, non-repudiation and integrity. The transform domain blind watermarking technique offers better visual & statistical imperceptibility and resiliency against different types of intentional & unintentional image degradations. Interleaving and interference cancellation methods are employed to improve the robustness performance significantly compared to conventional matched filter detection.

**Keywords**-Spread Spectrum Watermarking, Cellular Automata, Block Interleaving, Interference cancellation

## I. INTRODUCTION

With the prevalence of Internet, more and more digital data can be accessed via the network. Internet users can use multimedia data without offering appropriate credits to the creators. This hinders copyright owners form sharing their work in World Wide Web. Digital watermarking technique is the solution to the copyright protection problem of digital media. This models the authentication problem as a SS digital communication problem where an auxiliary message is embedded in multimedia signals and is available wherever the later signals move. The decoded message later on serves the purpose of copyright protection, data authentication, signal tagging, broadcast monitoring, security in communication etc. The art of the digital watermarking involves the judicious selection of technological trade-offs for different factors. These include cryptographic security, psychology of perception, high data embedding rate, robustness of extraction, statistical false extraction rates and complexity. All these requirements are related in conflicting manner and the particular algorithmic development emphasizes to a greater extent on few such requirements depending on the type of applications [1-4].

The SS watermarking provides an adequate cryptographic solution to the authentication of digital media. This paper addresses the implementation of watermark-authentication algorithm on CA based architecture. The objective of the present work stems from finding out such fact that have greater impact on improvement in detection reliability and authentication of SS watermarking scheme. In our views these depend on high quality CA pattern generator having appropriate rule selector for data authenticity and security. The local neighborhood additive rules provide pseudo exhaustive generators for cryptographic security. Here, watermark authentication function based on hybrid 1-D CA is proposed.

The paper is organized as follows: Section II introduces proposed watermarking technique within CA framework for image authentication. Section III discusses about the different features of the proposed scheme. Watermarking architecture is present in Section IV. Section V shows the implementation results on 16-by-16 binary watermark and finally section VI concludes and remarks about some of the aspects analyzed in this paper.

## II. PROPOSED SS WATERMARKING TECHNIQUE

Spread Spectrum watermarking is accomplished by embedding each watermark bit/ symbol over many samples of the cover image using a modulated pseudo random spreading sequence. SS watermark embedding and detection are analyzed mathematically in the following subsections.

### A. SS Watermark Insertion

(i) Let $B$ denotes the binary valued message bits string as a sequence of $N$ bits.

$$B = \{b_1, b_2, b_3, \ldots, b_n\}, b_i \; \varepsilon \; \{1, 0\} \quad (1)$$

(ii) Use the Two Predecessor Single Attractor (TPSA), Cellular Automata (CA) based scheme to perform the function of authentication of the message by the digital signature [5]. This digital signature is a function of shared secret key and message. The tags (i.e. digital signature) are mathematically represented as follows:

$$T_i = \{t_1, t_2, t_3, \ldots, t_n\}, t_i \; \varepsilon \; \{1, 0\} \quad (2)$$
$$i = 1, 2, 3, \ldots k$$

(iii) $M$–ary ($M = 2^m$) modulation map is used to produce modified digital signature of $M$–ary symbols [6] and can be expressed as follows:

$$T_i' = \{t_1', t_2', t_3', \ldots, t_n'\}, t_i \; \varepsilon \; \{1, 0\} \quad (3)$$

(iv) Block interleaving discussed in section III (B) is applied over every modified tag, and the resulting interleaved sequences are determined as shown below:

$$T_i'' = \{t_1'', t_2'', t_3'', \ldots, t_n''\} \quad (4)$$

where $T_i''$ is any one of $M$ possible number of bits per symbol ($t_1''$)

(v) Let $I$ denotes the image block of length $L$ i.e. image transform coefficients of length $L$, a bi–level (i.e. binary

valued) code pattern (discussed in section III) of length $L$ is used to spread each interleaved modified tag symbol. Thus $M$ different sets of code pattern $(U)$, each having $H$ number of bi–level modulation functions (since total number of different symbols in interleaved modified tag is $P$) of length $L$ are generated to form watermark sequence $W_L$.

$$[W_L] = \sum_{=1}^{L} (t_i'')_M \times [(U)_M], \quad (5)$$

$M = 2^m$ and $(U_M)_i$ is the $M$ th code pattern at $i$ th position.

(vi) The watermarked image $I_w$ can be obtained by embedding watermark information $W$ into the image block $I$. The data embedding operation can be mathematically expressed as follows:

$$[(I_W)_L] = [I_L] \pm \alpha \cdot [W_L] \quad (6)$$

where $\alpha$ is the modulation index which optimize or trade off maximum amount of allowable distortion (robustness) and minimum watermark energy (imperceptibility/ transparency) needed for a reliable watermark detection. $\alpha$ may or may not be a function of image coefficients. Accordingly SS watermarking schemes can be called as signal adaptive or non-adaptive SS watermarking. Watermark embedding strength can be determined from structure comparison $S(A; B)$ using cross covariance $\sigma_{ab}$ between cover $(A)$ and watermarked image $(B)$.

$$\sigma_{ab} = S(a,b) = \frac{1}{N-1} \sum_{i=1}^{N} (a_i - \mu_a)(b_i - \mu_b)$$

The structure comparison function after data embedding can be written as follows:

$$S(a,b) = (1/(N-1)) \left[ \sum_{i=1}^{N} (a_i - \mu_a)^2 + \sum_{i=1}^{N} (a_i - \mu_a)(\Delta a - \mu_{\Delta a}) \right]$$

$(\Delta a - \mu_{\Delta a})$ is small for high variance image coefficient block than low variance one in order to preserve $S(a, b)$ value same in both cases.

*B. SS Watermark Detection*

In SS watermarking, the detection reliability for binary valued message or data depends on decision variable $d_i$ obtained by evaluating zero lag spatial cross-covariance function between the watermarked image $I_w$ and each code pattern $[U_M]_H$. The decision variable $d_i$ can be mathematically expressed as follows:

$$d_i = \langle (U_M)_j - m_1((U_M)_j, I_W - m_1(I_W) \rangle (0) \quad (7)$$

where $m_1(S)$ represents the average of the sequence $S$. If $S_k$ represents the elements of $S$ with $k=1,2,3,...,L$ then $m_1(S)$ can be mathematically expressed as follows:

$$m_1(S) = \frac{1}{L} \sum_{k=1}^{L} s_k \quad (8)$$

The symbol (0) in equation (7) indicates the zero lag cross-correlation and for two sequences $S$ and $R$, the zero lag cross-correlation is given by,

$$\langle S, R \rangle (0) = \frac{1}{L} \sum_{k=1}^{L} s_k r_k \quad (9)$$

where $s_k$ and $r_k$ are the elements of sequence $S$ and $R$ respectively with $k=1,2,3,...,L$. The $M$ th symbol will be detected at the $i$ th position if the $M$ th cross correlation seems to be maximum than $(M-1)$ cross correlation indices at that position.. If the code pattern $(U_M)_i$ are chosen in such a way so that $m_1((U_M)_j) = 0 \ \forall j$, the computation of $d_i$ becomes

$$d_i = \left\langle (U_M)_j, \left[ I \pm \alpha \cdot \sum_{i=1}^{H} (t_i'')_M (U_M)_i \right] - m_1(I) \right\rangle$$

$$= \langle (U_M)_j, I \rangle \pm \alpha \cdot \sum_{i=1}^{H} (t_i'')_M \langle (U_M)_j, (U_M)_i \rangle - \langle (U_M)_j, m_1(I) \rangle$$

$$= \langle (U_M)_j, I_W \rangle \quad (10)$$

The above analysis indicates that code patterns used for spread spectrum watermarking should posses some specific properties. Watermark detection is improved if the following conditions are satisfied.

$(U_M)_j, j = 1, 2, \ldots L$ should be distinct sequences with zero average.

The spatial correlation $= \langle (U_M)_i, (U_M)_j \rangle$ $i \neq j$ should be minimized. Ideally, sequences $(U_M)_i$ and $(U_M)_j$ should be orthogonal.

Each $(U_M)_j$ for $j = 1, 2, \ldots, L$ should be uncorrelated with image block $I$ when image prediction (for estimating the image distortion) is not used before evaluating the cross–correlation.

III. WATERMARKING PRELIMINARIES

*A. Cellular Automata*

It represents the sequential behavior of a number of interconnected cells arranged in regular manner. Here, next state of each cell depends on the present state of its neighbor cells [7]. If a three neighborhood is considered, the state $q$ of $i$-th cell at time $(t+1)$ can be denoted as $q_i^{(t+1)} = g(q_{i-1}^t, q_i^t, q_{i+1}^t)$ where $q_i^t$ denotes the state of the $i$-th cell at time $t$ and $g$ is the next state function. For $n$ cell one-dimensional CA, the linear operator is a $(n \times n)$ matrix called as characteristic matrix $(T)$. If $f_t$ represents the state of the CA at the $i$-th instant of time, then the following relation gives the next state of the CA

$$f(t+1) = T \times f_t \text{ and thus } f_{(t+p)} = T^p \times f_t$$

For linear one dimensional cellular automata (CA), the next state function for Rule 90 and Rule 150 are given below:

Rule 90: $q_i^{(t+1)} = q_{i-1}^t \oplus q_{i+1}^t$

Rule 150: $q_i^{(t+1)} = q_{i-1}^t \oplus q_i^t \oplus q_{i+1}^t$

*B. Block Interleaving*

In digital watermarking, the signal detector has accessed to all the encoded data at once and the signal layout is in *2-D*. Here, errors occur in bursts that stretch across the horizontal and vertical axes corrupting a cluster of symbols in the array. For communication theory based watermarking application and using linear codes, a well-known technique (Block Interleaving) is used to overcome bursts of errors [8]. Let there exists a *1-D* block of length $l^2$ symbols. Therefore, interleaving can be obtained by using a *2-D* memory array of size *(l x l)* symbols having interleaving degree $\lambda = l$. Symbols are written to memory column-by-column and read row-by-row (or vice versa). If the linear code used is capable of correcting a single bursts error of length *t* symbols or less in a block of length *l*, the interleaved code will correct any single bursts of length $\lambda l$ or less.

C. *Transform for Data Hiding*

It provides co-joint representation in the form of space spatial frequency resolution with local and global information of the image signal.

(1) DWT: DWT system decomposes an image signal using orthogonal filter (e.g *db2*) into four *LL* (approximation), *LH* (horizontal), *HL* (vertical) and *HH* (diagonal) sub bands, thereby giving logarithmic resolution [9].

(2) *M*-Band WT: This system decomposes the image into (*M* X *M*) channels corresponding to different directions and resolution resulting in linear and logarithmic resolution [10].

(3) Bi-orthogonal Wavelet: In case of DWT and *M*-Band WT, imperceptibility and robustness performance become lower with increasing embedding rate. Interference occurs as code patterns are added to decomposed image using single scaling function. Moreover, this doesn't yield low correlation between image and spreading sequence, therefore high robustness may not be achieved. This problem can be solved using Bi-orthogonal DWT (Bi-DWT), which provides low correlation with code patterns [11].

(4) Hilbert Transform: Similar to Bi-DWT, several authors have proposed signal processing methods that call for two wavelet transforms where one wavelet is (approximately) the Hilbert transform of the other [12]. The definition of Hilbert transform is mathematically expressed as follows: $\psi_g(t)$ is the Hilbert transform of $\psi_h(t)$ if

$$\Psi_g(\omega) = -j\ \Psi_h(\omega),\ \omega > 0$$
$$= j\ \Psi_h(\omega),\ \omega < 0$$

D. *Complementary Modulation*

For the random modulation techniques there are four possible types of modulations: *MF (+, +), MF (+, -), MF (-, +)* and *MF (-, -)* where *MF* stands for modulation function. *MF (+/-, -/+)* represents a positive/ negative transformed coefficient modulated with a negative/ positive watermark quantity [13]. The influence of different attacks is also observed to see how they update the magnitude of transformed coefficients. These attacks can be roughly classified into two categories. The first category (like blurring, compression etc.) decreases the magnitude of most of the transformed coefficients. So, every coefficient can be modulated with *MF (+, -)* and *MF (-, +),* so that they will contribute positively to the detector response. On the other hand, second category (like sharpening, histogram equalization) tends to increase the coefficient magnitude; thereby *MF (+, +)* and *MF (-, -)* will similarly improve the detector response. On the basis of this analysis, complementary modulation strategy is presented. The proposed scheme embeds two identical watermarks, which play complementary roles in resisting various kinds of attacks. One watermark is embedded using negative modulation rule with *MF (+, -)* and *MF (-, +)*. The second watermark follows the positive modulation rule using *MF (+, +)* and *MF (-, -)*.

E. *Spreading Code Pattern*

High detection reliability is achieved if the code pattern sequences posses very low zero lag cross-correlation among each other as well as with image block when image prediction is not used to evaluate the cross correlation. The code patterns used for SS modulation are pseudo-random or pseudo-noise (PN) sequences. These sequences can be generated by linear congruential generators. But the desired property of $m_I[\langle (U_M)_j]= 0$ and $\langle (U_M)_i, (U_M)_j \rangle = 0$ for i≠j can be theoretically guaranteed only for infinite length sequence which is not feasible in practice. Optimal small set Kasami sequence can improve the detection reliability by matching Welch's lower bound with good cross correlation properties. Gold sequence also shows good periodic cross correlation properties and also provides better detection [14]. Mayer *et al.* proposed deterministic sequence generation from Walsh/Hadamard basis and Gram-Schmidt orthogonalisation of PN sequences. In order to remove the deterministic nature of the basis function for authenticity, Walsh/Hadamard basis is used to modulate the spreading code patterns. This is analogous to Walsh covering in DS-CDMA digital cellular system (IS-95) by Qual Comm Inc. Gram-Schmidt orthogonalised sequence is normalized so that on an average same distortion is introduced and techniques may be fairly compared.

IV. PERFORMANCE ANALYSIS & EVALUATION

We observe the error performance of the proposed watermarking scheme. Nodes other than source and destination may also receive data and it is such that no nodes except destination node can be able to detect the original data. Hence authenticity, security and integrity of embedded data is maintained by additive neighborhood CA rules. Relative entropy distance (Kullback Leibler distance) is also used as a measure of security. Here, peak signal to noise ratio (PSNR) and structural similarity index measurement (SSIM) are used as representative objective measure of data imperceptibility. In decoder, symbol-by-symbol of hard decoding is considered assuming magnitude of interference between host signal and code pattern is much smaller than interference due to multiple symbol hiding. Mathematically the symbol error probability can be obtained as follows:

$P_{sym} = (1/N)[P\ (\varepsilon\ /\ Y^{sym1}) + P\ (\varepsilon\ /\ Y^{sym2}) + ... + P\ (\varepsilon\ /\ Y^{symN})]$

where ε indicates the error. It is the sum of non-gaussian random variables. If the number of terms are large enough, central limit theorem may be applied and approximated by a gaussian distribution with mean (μ) and variance (σ$^r$) calculated as follows:

$X_k = \langle I_l', G_{kl} \rangle = \Sigma\Sigma(I_l + \alpha.G_{il}).G_{kl};\ \mu_i = (1/N)\Sigma X_i$
$\quad i=1:N\quad l=1:L \quad\quad\quad i=1:\ N$

$\sigma_i^2 = var\ [X_k]$, L indicates number of signal points. The conditional distribution of detector statistics is given by:

$f(x_i/g_i) \cong (1/(2\pi\sigma_i^2)^{0.5})\ exp[(x_I - \mu_i)/2\sigma_i^2]$

The robustness performance of proposed technique is evaluated for various signal processing operations e.g. linear-nonlinear filtering, JPEG & JPEG-2000 compressions etc.

over large number of benchmark images. The proposed algorithm shows greater resiliency against above stated image impairments since each symbol is embedded in two different sub bands using complimentary modulation functions. Redundancy in data hiding offers better stability for overall mean cross-correlation value and probability of error becomes low even after high degree of signal degradation. Moreover, robustness performance of Bi DWT and Hilbert transform pair of WT are much better than DWT as former decomposition offers low correlation with code patterns and higher energy content for *HH* sub bands compared to later decomposition. Robustness performance of *M*–ary modulation is better for *M*–Band wavelet decomposition compared to DWT with the increased value of *m*. Here computational cost of decoding is also increased as embedded channel is projected onto all modulation functions of that particular position for each set of key. We test the performance improvement of detection process using block interleaving along with serial and parallel interference cancellation (SIC & PIC). This improves the capacity of watermarking technique further. For SIC, decision satisfies for an embedded symbol is obtained by subtracting an estimate of already detected symbols from received signals. For PIC, decision statistics is determined for all symbols parallely in the same way until two consecutive steps are become identical. The performance of PIC is much better than SIC and block-interleaving technique improves the subjective as well as objective measurement of embedded information. Table I and II show the implementation results of the proposed scheme.

## V. CONCLUSION

Cellular automata based spread spectrum image watermarking scheme is critically analyzed in wavelet transform domain for multiple message embedding using complimentary modulation function with *M*–ary modulation. Experimental results show that data security, embedding rate can be significantly improved. Moreover, block-interleaving technique along with SIC/PIC improves robustness-capacity to a greater extent compared to conventional correlation based approach.

## ACKNOWLEDGMENT

Dr. Subir Kumar Sarkar acknowledges the financial assistance obtained from the Centre for Mobile Computing and Communication, Jadavpur University, Kolkata.

## REFERENCES

[1] H. S. Malvar and D. A. F Florencio, "Improved Spread Spectrum: A new Modulation Technique for Robust Watermarking," *IEEE Trans on signal processing*, pp 898-905, April 2003.
[2] I. J. Cox, J. Killian, F. T. Leighton, T. Shamoon, "Secured Spread Spectrum Watermarking for Multimedia," *IEEE Trans on Image Processing*, vol. 6, pp 601-610, June 2000.
[3] R. Grobois, T. Ebrahim, "Watermarking in JPEG-2000 Domain," *Proc. of IEEE Workshop on Multimedia Signal Processing,* pp. 3-5, 2001.
[4] J. Mayer, A. V. Silverio, J. C. M. Bermudez, "On the Design of Pattern Sequences for Spread Spectrum Image Watermarking," *International Telecommunications Symposium,* Brazil.
[5] P. Dasgupta, S. Chattopadyay, I. Sengupta, "An ASIC for Cellular Automata based Message Authentication," In Proc. Intl. Conf. on VLSI Design, India, pages 538-541, January 1999.
[6] S. P. Maity, M. K. Kundu, T. S. Das, P. K. Nandi, "Robustness Improvement in Spread Spectrum Watermarking using M-ary Modulation," *Proc. of Nat. conf. on Communication*, pp. 569-573, Jan 2005.
[7] S. K. Sarkar, *et.al.*, "A One Dimensional Cellular Automata Based Security Scheme Providing Both Authentication and Confidentiality," *Journal of IEE* , Vol. 87, pp. 1-7, May 2006.
[8] Peterson and Weldon, "Error Correcting Codes," 2[nd] Ed. *MIT Presss Cambridge, MA*, 1972.
[9] I. Daubechis, "Orthogonal bases for Compactly Supported Wavelets", *Comm., Pure Appl. Math*, Vol. 41, pp 909-996, 1988.
[10] M.K.Kundu and M.Acharya, "M-Band Wavelets: Application to Texture Segmentation for Real Life Image Analysis," *International journal of wavelets, Multiresolution and Information proceeding*, vol 1, pp115-149, 2003.
[11] C. S. Burrus, R. A. Gopinath, H. Guo, "Introduction to Wavelets and Wavelet Transform," *A Primer, Prentice Hall, NJ*, 1997
[12] I. Selesniek, "Hilbert Transform Pairs of Wavelet Bases," *IEEE Signal Processing Letters* Vol. 8, No.6, 2001.
[13] C-S. Lu, S-K. Huang, C-J. Sze, H-Y. M. Liao, "Cocktail Watermarking for Digital Image Protection," *IEEE Transaction on Multimedia* Vol. 2, No. 4, 2000.
[14] D. V. Sarwate, M. B. Purseley, "Correlation Properties of Pseudorandom and Related Sequences," *Proceedings of IEEE* Vol. 68, No. 5, pp. 583-619, 1980.

TABLE I
NUMERICAL RESULTS

| Image | Filter | SSIM | PSNR | Security Value |
|---|---|---|---|---|
| Fishing Boat | DWT/ db2 | 0.973 | 38.318 | 0.0185 |
| | M Band | 0.985 | 40.041 | 0.0159 |
| | Bi-DWT | 0.989 | 40.423 | 0.0160 |
| | Hilbert Transform | 0.098 | 39.326 | 0.0162 |
| Lena | DWT/ db2 | 0.970 | 38.307 | 0.0187 |
| | M Band | 0.983 | 39.551 | 0.0160 |
| | Bi-DWT | 0.987 | 39.973 | 0.0160 |
| | Hilbert Transform | 0.976 | 39.303 | 0.0163 |

TABLE II
JPEG-2000 COMPRESSION

| QF | M = 2 $P_e$ | | M = 4 $P_e$ | | M = 8 $P_e$ | | M = 16 $P_e$ | |
|---|---|---|---|---|---|---|---|---|
| Q | SIC | PIC | SIC | PIC | SIC | PIC | SIC | PIC |
| 100 | 0.1 | 0.085 | 0.0 | 0.0 | 0.0 | 0.0 | 0.0 | 0.0 |
| 75 | 0.3 | 0.26 | 0.1 | 0.073 | 0.1 | 0.089 | 0.057 | 0.017 |
| 50 | 0.4 | 0.38 | 0.3 | 0.262 | 0.2 | 0.179 | 0.15 | 0.103 |
| 25 | 0.5 | 0.47 | 0.4 | 0.312 | 0.3 | 0.237 | 0.3 | 0.235 |
| 0 | 0.6 | 0.53 | 0.45 | 0.353 | 0.4 | 0.320 | 0.35 | 0.317 |